\documentclass[prc,aps,showpacs,twocolumn]{revtex4}
\usepackage{bm}
\usepackage{dcolumn}
\usepackage{graphicx}
\usepackage{amsmath}
\usepackage{amssymb}
\usepackage{color}
\usepackage{bbold}
\newcommand{\beq}{\begin{equation}}
\newcommand{\eeq}{\end{equation}}
\newcommand{\beqa}{\begin{eqnarray}}
\newcommand{\eeqa}{\end{eqnarray}}

\newcommand{\bs}[1]{\ensuremath{\boldsymbol{#1}}}
\definecolor{green2}{rgb}{0,0.6,0.1}

\begin{document}
\title{
{Role of final state interaction and of three-body force\\ 
on the longitudinal response function of $^4$He}}
\author{Sonia Bacca$^{a}$,
Nir Barnea$^{b}$, Winfried Leidemann$^{c}$
and Giuseppina Orlandini$^{c}$}

\affiliation{
$^{a}$TRIUMF, 4004 Wesbrook Mall,
Vancouver, B.C. V6J 2A3, Canada\\
$^{b}$Racah Institute of Physics, Hebrew University, 91904, Jerusalem,
Israel\\ 
$^{c}$Dipartimento di Fisica, Universit\`{a} di Trento and INFN\\
(Gruppo Collegato di Trento), via Sommarive 14, I-38100 Trento, Italy}
\date{\today}
\begin{abstract}
We present an {\it ab-initio} calculation of the longitudinal electron scattering response function off $^4$He with two- and three-nucleon forces and compare to experimental data. The full four-body continuum dynamics is considered via the Lorentz integral transform  method. The importance of the final state interaction is shown at various energies and momentum transfers $q$. The three-nucleon force reduces the quasi-elastic peak by 10\% for $q$  between 300 and 500 MeV/c. Its effect increases significantly at lower $q$, up to about 40\% at $q$=100 MeV/c. At very low $q$, however, data are missing.  
\end{abstract}

\pacs{25.30.Fj, 21.45.-v, 27.10.+h, 31.15.xj}
\maketitle

Inelastic electron scattering off nuclei provides important informations on  nuclear dynamics.  
Varying the momentum $q$, transferred by the electron to the nucleus, one
can focus on different  
dynamical regimes. At lower $q$  the collective behavior of nucleons is studied. As  $q$  
increases one probes  properties of the  single nucleon in the nuclear medium 
and its correlations to other nucleons from long- to short-range.
Thus the {\it inclusive} longitudinal $R_L$
and transverse $R_T$
response functions 
are of particular importance.
Different from  $R_T$,
 in a non-relativistic framework $R_L$ does not require the knowledge of implicit degrees of freedom (exchange currents),
providing a clean
leptonic probe of the nuclear Hamiltonian.
In addition, the
theoretical study of {\it inclusive} processes is important to help planning further investigations, 
for selected kinematics, via  {\it exclusive} scattering experiments.

In the '80 and '90's an intense experimental activity has been devoted to  {\it inclusive} electron
scattering, $(e,e')$,  in the so called quasi-elastic (q.e.) regime, corresponding to $q$-values 
of several hundred MeV/c and energy transfers $\omega$ around the q.e. peak ($\omega \simeq q^2/2m$). 
Here one can  envisage that the electron has scattered elastically with a single nucleon of 
mass $m$. Various nuclear targets have been considered, from very light 
to heavy ones 
~\cite{reportexp}. 
At these $q$ one enters a very challenging regime, where nuclear and subnuclear 
degrees of freedom interwine.
A very alive debate has taken place about the interpretation of those data.
The two most discussed topics 
have been:
$(i)$ short-range correlations, {\it i.e.} the 
dynamical properties of nucleons at short distances;
$(ii)$ in medium modifications of 
the nucleon form factor. To date the debate is still open.
More experiments are planned at 
Jefferson Laboratory (E05.110 at Hall A) which will contribute to those issues  and a 
theoretical effort is needed to help interpreting old and new experimental results. 

The reason for concentrating on the q.e. regime has been the conviction that for such a kinematics  
 the plane wave impulse approximation (PWIA) might 
be a reliable framework to describe the reaction.
The neglect of the final state interaction (FSI) has the advantage to allow a simple interpretation 
of the cross section in terms of the dynamical properties of the nucleons in the ground state. Thus it is important to clarify
the reliability of the PWIA (as well as of further refinements). The Euclidean   
approach ~\cite{Euclidean} has already shown that the PWIA is rather poor, however,
this method does not easily allow to obtain the $\omega$-dependence of the FSI effects. 

The aim of this letter is twofold. On the one hand we study the role of FSI on
$R_L$ of $^4$He at 300 MeV/c $\le$ $q$ $\le$ 500 MeV/c, where by now only calculations with central
two-nucleon forces exist \cite{ELO97,Sonia07}. Here we use a realistic two-body potential
augmented by a three-nucleon force (3NF)  and compare the PWIA  to results obtained via the Lorentz integral transform (LIT) 
method~\cite{EFROS94,REPORT07}. 
The LIT method is an {\it ab-initio} approach, which allows the full treatment of the four-body problem.
It has already been applied to various realistic calculations of electroweak reactions in
three- \cite{ELOT2000,Golak2002,ELOT2004,ELOT2005,Sara2008} and four-body systems \cite{Doron2006,Doron2007}.
Different from the Euclidean approach, the LIT method allows a comparison
with the PWIA regarding the $\omega$-dependence of $R_L$. Our second focus lies
on the study of the role of 3NFs. We 
contribute to this much debated issue investigating  3NF effects on initial and
final states by studying $R_L$ in various kinematical regions.

The choice of $^4$He as a target is of particular interest. In fact 
$^4$He has quite a large average density. Moreover its binding energy per 
particle is similar to that of heavier systems. Therefore $^4$He results can 
serve better as guidelines for investigating  heavier nuclei than results for two- 
and three-body systems. Various {\it inclusive} $^4$He $(e,e')$ experiments 
have been performed in the past (see~\cite{SICK}  for a summary of the world data), and a 
comparison theory-experiment is possible without the ambiguities, created by the Coulomb 
distortions, which affect heavier systems. 

The longitudinal response function  is given by
\beq
\nonumber
R_L(\omega,q)=\int \!\!\!\!\!\!\!\sum _{f} 
|\left\langle \Psi_{f}| \hat\rho(q)| 
\Psi _{0}\right\rangle|
^{2}\delta\left(E_{f}+\frac{q^2}{2M}-E_{0}-\omega \right), 
\eeq
where  
$M$ is the target mass, $| \Psi_{0/f} \rangle$ and 
$E_{0/f}$ denote initial and final state wave functions and energies,
respectively.
The charge density operator $\hat{\rho}$ is defined as
\beq
{\hat\rho}(q)= \frac{e}{2} \sum_i \,(1 + \tau_i^3) \exp{[i {\bf q} \cdot {\bf r}_i]} \,, 
\eeq
where $e$ is the proton charge and $\tau_i^3$ the isospin third component of nucleon $i$.
The $\delta$-function 
ensures energy conservation. $R_L$ contains a sum over
all  possible final states, which are excited by the electromagnetic
probe, including also continuum states. Thus, in a straightforward evaluation one 
would need to calculate both 
bound and continuum states. The latter constitute the major
obstacle for  many-body systems if one wants to treat the nuclear
interaction rigorously. In the LIT method~\cite{EFROS94,REPORT07} this difficulty is
circumvented  by considering instead of 
$R_L(\omega,q)$ an integral transform 
${\cal L}_L(\sigma,q)$ with a Lorentzian kernel defined for a complex
parameter $\sigma=\sigma_R+i\,\sigma_I$ by
\begin{equation}
  {\cal L}_L(\sigma,q)=\int d\omega
  \frac{R_L(\omega,q)}
  {(\omega-\sigma_R)^{2}+\sigma_I^{2}}
  = \langle\widetilde{\Psi}_{\sigma,q}^{\rho}
  |\widetilde{\Psi}_{\sigma,q}^{\rho}\rangle \,.\label{lorentz_transform} 
\end{equation}
The parameter $\sigma_I$ determines the resolution of ${\cal L}_L$ and is kept at a constant 
finite value ($\sigma_I\ne 0$). The basic idea of considering ${\cal L}_L$
lies in the fact that it can be evaluated from the norm of a
function $\widetilde{\Psi}_{\sigma,q}^{\rho}$,
which is the unique solution of the inhomogeneous equation 
\begin{equation}
(\hat{H}-E_{0}-\sigma)|\widetilde{\Psi}_{\sigma,q}^{\rho}
\rangle=\hat{\rho}(q)|{\Psi_{0}}\rangle\,.\label{liteq}
\end{equation}
Here $H$ denotes the nuclear Hamiltonian.
Due to the presence of the imaginary part $\sigma_I$ in (\ref{liteq}) 
and the fact that its right-hand side is
localized, one has a bound-state like asymptotic boundary condition. 
Thus, one can apply bound-state techniques for its
solution. 
Finally, $R_L(\omega,q=const)$ is obtained by inverting
the LIT (\ref{lorentz_transform}). Subsequently the isoscalar 
and isovector parts of $R_L$ are multiplied by the proper nucleon form factors.
For the LIT inversion various
methods have been devised~\cite{EfL99,AnL05}. 
\begin{figure}
\includegraphics[scale=0.29,clip=]{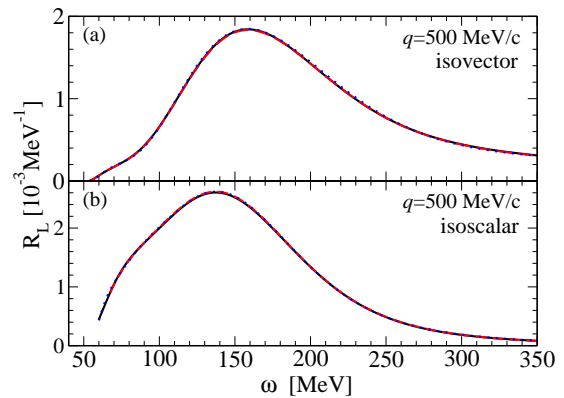}
\vspace*{-3mm}
\caption{Isovector (a) and isoscalar (b) parts of $R_L(\omega,q)$ at $q$=500 MeV/c. 
Single multipole contributions with $K_{max}^{JT}$ (solid) and with $K_{max}^{JT}-2$ (dashed)
first inverted and then summed up; single multipole contributions with  $K_{max}^{JT}$ first summed up and then  inverted (dotted). 
} 
\label{FIG1}
\end{figure}

The PWIA result is obtained under the hypothesis of one outgoing free proton with mass $m$ and a spectator (A-1)-system with mass $M_s$:
\begin{equation} 
  R_L^{PWIA}(\omega,q)=\! \int d{\boldsymbol p}  \,n({\boldsymbol p})\, 
  \delta \left( \omega-\frac{( {\boldsymbol p} + {\boldsymbol q} )^2}{2
  m}-\frac{\bs{p}^2}{2 M_{s}} - \epsilon \right). 
  \nonumber
\end{equation}
Here $n({\boldsymbol p})$ represents the proton momentum distribution and
$\epsilon$ the proton separation energy.
In the following we present results obtained with the Argonne V18 (AV18)
\cite{AV18} and the Urbana IX (UIX)~\cite{UIX} two- and three-body forces.
As nucleon form factor we use the proton dipole fit and the neutron electric form factor from 
\cite{dipole}.
The solution of (\ref{liteq}), as well as the ground state $|{\Psi_{0}}\rangle$, 
is expanded in hyperspherical harmonics (HH). The HH expansion is truncated
beyond a maximum value $K_{max}$ of the HH grand-angular
momentum quantum number.  The convergence of the HH expansion is
improved by introducing a $K_{max}$-dependent effective
interaction (EIHH-method)~\cite{BaL00,EIHH-3NF}.
In order to evaluate ${\cal L}_L$ we have calculated the norm
$\langle\widetilde{\Psi}_{\sigma,q}^{\rho}
|\widetilde{\Psi}_{\sigma,q}^{\rho}\rangle$ directly,
using the Lanczos algorithm~\cite{marchisio}. 
The operator $\hat\rho$ is expanded in Coulomb multipoles of order $J$.
The LIT is calculated for each isoscalar ($T$=0) and isovector ($T$=1) multipole separately up 
to a maximal value of $J_{max}$  where 
convergence of the expansion is reached. The values of $J_{max}$ vary from 2 to 7
for $q$ ranging from 50 to 500 MeV/c.

The accuracy of the results is determined mainly by the convergence of the HH expansion
and the stability of the inversion. 
In the calculations we used a ground state hyperspherical momentum value  $K^0_{max}=$16 (14) for the AV18+UIX (AV18) case, leading to a binding energy of  28.4 (24.3) MeV.
Since a multipole dependent convergence pattern  has been encountered and each multipole contributes 
differently to the total strength,  
the  $K_{max}$ used for the LIT evaluation vary according to the value of $J$, namely   $K_{max}^{JT}= 12 - 16$  for even $J$ and  $K_{max}^{JT}=13-17$  for the odd $J$ have been considered.
Our LIT results converge at a percentage level.
In Fig.~\ref{FIG1} the accuracy of the results for  $R_L$ regarding both the HH expansion and 
the inversion stability aspects is illustrated exemplary
for the isoscalar and isovector parts at $q$=500 MeV/c. The figures contain three curves: 
the full line is obtained when
 the single multipole contributions ${\cal L}_L^{JT}$, calculated up to $K_{max}^{JT}$, are first inverted and then summed up.
The dashed line represents the results where the various
multipole contributions ${\cal L}_L^{JT}$ are calculated only up to $K_{max}^{JT}-2$. 
The comparison between these two results illustrates the quality of the HH convergence.
The dotted line reflects  the inversion of the total ${\cal L}_L(\sigma,q)$, where the various
multipole contributions ${\cal L}_L^{JT}$, calculated up to $K_{max}^{JT}$, are first summed up and then inverted. 
The comparison between the dotted and full lines shows the accuracy 
of the inversion.
\begin{figure}
\includegraphics[scale=0.38,clip=]{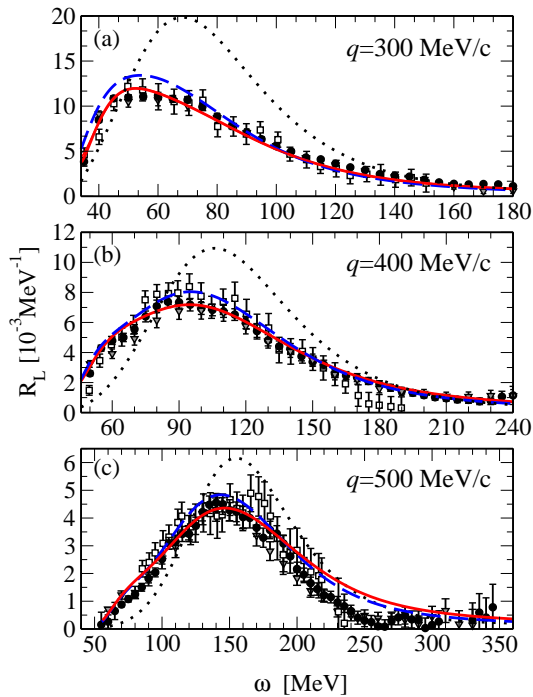}
\vspace*{-3mm}
\caption{$R_L(\omega,q)$ at various $q$: PWIA using $n({\bf p})$ of AV18+UIX~\cite{Wiringa} (dotted); full calculation with AV18 (dashed) and AV18+UIX
(solid). Data from Bates~\cite{Bates} (squares), Saclay~\cite{Saclay} (circles) and world-data set from \cite{SICK} (triangles). 
} 
\label{FIG2}
\end{figure}
In Fig.~1 one finds very satisfying results for both isospin channels for
the HH convergence and the accuracy of the inversion as well. We should mention
that we do not show the low-energy isoscalar response, where a narrow $0^+$ resonance
with a width of a few hundred keV is present at $E_r$ very close to threshold \cite{Wal70}. 
To get accurate results for such a resonance a convergent LIT calculation with 
a $\sigma_I$ much smaller than the presently used values  (smallest value $\sigma_I=5$  MeV)
 should be carried out, which then leads to a very slow
asymptotically fall off of the solution $|\widetilde{\Psi}_{\sigma,q}^{\rho}\rangle$
(see~\cite{winfried08}). Such a calculation requires a considerable additional computational
effort and thus the threshold region is excluded from our present work.
Allowing a narrow resonance in the inversion \cite{winfried08}, we have checked
that our results are stable for energies above $E_{r}+2\sigma_I$.

\begin{figure}
\includegraphics[scale=0.38,clip=]{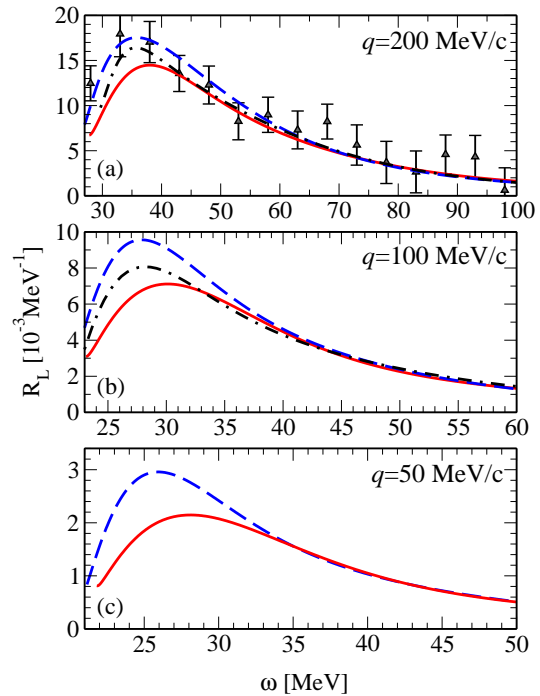}
\vspace*{-3mm}
\caption{$R_L(\omega,q)$ at various $q$ with the  AV18 (dashed),
 AV18+UIX (solid) and MTI-III (dash-dotted) potential. 
Data in (a) from~\cite{Buki}.
} 
\label{FIG3}
\end{figure}

\begin{table}
\caption{$R_L$ peak position $\omega_p$ and $R_L$ peak height without
3NF (AV18),
with 3NF (AV18+UIX), and relative 3NF effect
$\Delta R = 100 \times (R_L({\rm AV18})-R_L({\rm AV18+UIX}))/R_L({\rm AV18})$.
\label{table1}
}
\begin{center}
\begin{tabular}
{cc cc cc cc cc cc}\hline\hline

               & AV18
               & AV18+UIX
               & AV18
               & AV18+UIX
               &        \\ \hline
   $q$           &  $\omega_p$
               &  $\omega_p$
               &  $R_L(\omega_p,q)$
               &  $R_L(\omega_p,q)$
               &  $\Delta R$   \\ \hline
 [MeV/c]  & [MeV]  & [MeV]  & [$10^{-3}$MeV$^{-1}$]  &
[$10^{-3}$MeV$^{-1}$]  & [\%]  \\

\hline\hline

50  &   26  &   28  &   2.96 &  2.15 &  -27  \\
100  &  28  &   30  &   9.56 &  7.11 &  -26  \\
200  &  36  &   38  &   17.5 &  14.5 &  -17  \\
300  &  54  &   52  &   13.4 &  12.0 &  -10  \\
350  &  73  &   70  &   10.3 &  9.20 &  -11  \\
400  &  95  &   95  &   8.04 &  7.18 &  -11  \\
500  & 143  &  146  &   4.84 &  4.36 &  -10  \\
\hline\hline
\end{tabular}
\end{center}
\end{table}
In Fig.~\ref{FIG2} the results of $R_L(\omega,q)$ at various $q$ are shown and
compared to data. In all cases one finds that the FSI effects are very large
and essential for reaching agreement with experiment. 
The PWIA fails particularly 
in the q.e. peak and at low $\omega$.
With growing $q$ FSI effects decrease in the peak region, but not at low $\omega$. 
One may also consider a more refined PWIA, where a spectral function is used
instead of a momentum distribution (see {\it e.g.}~\cite{SpectralFnostra}). In~\cite{SpectralFnostra} it was 
shown that such an improved PWIA  modifies the  simple PWIA result by only 10-20 \%.

In Fig.~\ref{FIG2} one also sees the 3NF effects on the full calculation. For $q$=300 MeV/c one notes a good agreement of the data with the AV18+UIX result.
This is true for $q$=400 MeV/c as well, if one does not consider the data of~\cite{Bates}, which exhibit larger error bars.
At $q$=500 MeV/c some discrepancies between theory and experiment are present in
the low- and high-energy range, while there is a fairly good agreement in the peak region. However, investigations on the three-body systems~\cite{ELOT2005} have shown that a 
 consideration of relativistic effects becomes important at such a momentum transfer.

Table~\ref{table1} illustrates  the 3NF effect 
on peak position and peak height also for lower $q$.
One notes that there is no unique 3NF effect on the position, while
one has a reduction of the height due to the 3NF at any $q$. The size of the reduction amounts to 10\%
for the higher $q$, whereas below $q$=300 MeV/c the reduction grows with 
decreasing $q$, reaching almost 30\% at $q\le$100 MeV/c.
In Fig.~\ref{FIG3} the results at lower $q$ are shown.
The important role of the 3NF is evident in the whole peak 
region, leading to a strong decrease of $R_L$ of up to 40\% for some $\omega$ values.
Recently also some new data at $q \simeq 200$ MeV/c have been published~\cite{Buki}
(see Fig.~\ref{FIG3}a). While one finds a satisfactory agreement between the AV18+UIX
result and data beyond the peak, one observes a non negligible discrepancy in the peak itself.
In Figs.~\ref{FIG3}a,~\ref{FIG3}b we also illustrate $R_L$ for a calculation
\cite{Sonia07} with a central  
two-nucleon potential (MTI/III model \cite{MaT69}). Results are more similar to the AV18
than to the AV18+UIX curves, showing that the 3NF effect is not  simply
  explained by the binding energy difference 
($^4$He binding energy with AV18, AV18+UIX, and MTI/III is 24.3, 28.4 and 30.6 MeV, 
respectively).

We summarize our results as follows. We have carried out an {\it ab-initio} calculation of the 
longitudinal $(e,e')$ response function $R_L(\omega,q)$  of $^4$He for various 
kinematics up to $q$=500 MeV/c. The full dynamics of the four-body system has been taken 
into account
for the $^4$He ground state and the four-body continuum states as well. The rigorous inclusion of FSI
has been achieved by use of the LIT method. Our work has been mainly focused on
two points, namely the study of the importance of FSI and of 3NF. We have shown that both 
ingredients play an important role and need to be considered in a calculation of $R_L$.
A particularly important finding are the very large 3NF effects of up to 40\% 
in the $R_L$ peak region at $q \le 200$ MeV/c. 
Thus it is becoming apparent that 
there exists an electromagnetic observable, complementary to the purely hadronic ones, where one can learn more about the not yet well established 3NF.
In view of our findings we 
hope for
 a revival of the experimental interest in electron 
scattering, especially on light nuclei and at lower energies and momenta.

The work of N. Barnea was supported by the Israel Science Foundation
(grant no.~361/05).
This work was supported in part by the
Natural Sciences and Engineering Research Council (NSERC) and by the
National Research Council of Canada.
Numerical calculations were  performed at CINECA (Bologna).

\end{document}